\documentclass[12pt]{article}
\usepackage{amstext}
\usepackage{amsmath}
\usepackage{graphicx}
\usepackage[latin1]{inputenc}
\usepackage{pstricks}
\usepackage{pst-plot}
\usepackage{soul}

\begin{document}

\date{}
\title{Subtleties on energy calculations in the image method}
\author{M.M. Taddei$^1$, T.N.C. Mendes$^2$,
C. Farina$^3$\\
$^{1,3}$Instituto de F\'{\i}sica - UFRJ - CP 68528\\
Rio de Janeiro, RJ, Brazil - 21945-970.\\
$^{2}$Escola de Ciências e Tecnologia,\\
Universidade Federal do Rio Grande de Norte, Natal, Brazil}
\maketitle

\begin{abstract}

In this pedagogical work we point out a subtle mistake that can be
done by undergraduate or graduate students in the computation of the
electrostatic energy of a system containing charges and perfect
conductors if they naively use the image method. Specifically, we
show that naive expressions for the electrostatic energy for
these systems obtained directly from the image method are wrong by a
factor $1/2$. We start our discussion with well known examples,
namely, point charge-perfectly conducting wall and point charge-perfectly
conducting sphere and then proceed to the demonstration of general
results, valid for conductors of arbitrary shapes.

\end{abstract}

\bigskip

 \vfill \noindent

 \noindent
  $^1$ {e-mail: marciotaddei@if.ufrj.br}

  \noindent
$^{2}$ {e-mail: tarciromendes@ect.ufrn.br}

\noindent $^3$ {e-mail: farina@if.ufrj.br}

\section{Introduction}

The typical problem in electrostatics consists of determining in all relevant space the electrostatic field generated by some set of charges from the known charge distribution itself as well as from the appropriate boundary conditions pertaining to the situation. Solving this problem amounts to finding a static potential $\Phi({\bf x})$ obeying Poisson's equation, namely,
\begin{equation}
\nabla^2\Phi({\bf x})=-\dfrac{\rho({\bf x})}{\epsilon_0} \ ,
\label{Poisson}
\end{equation}
subject to the suitable boundary conditions. Although one can approach this task in many different ways -- whose convenience depends on the particular problem being dealt with --, Poisson's equation (given definite boundary conditions) has a unique solution for each charge distribution $\rho({\bf x})$. This allows one to look for solutions in any desired fashion: if one finds a potential that obeys both the boundary conditions and Poisson's equation for the correct $\rho({\bf x})$, it must be the correct potential for the given configuration.

A especially suitable method for simple situations with point charges and dipoles in the presence of conductors is called the image method. It consists of finding a different configuration in which the conductors are replaced by some charge distribution so that the potential of the entire setup in the region of physical interest (i.e., outside the conductors) fits the appropriate boundary conditions and also obeys Poisson's equation. These fictitious charges put  in place of the conductors are called image charges. Since any potential created by charges obeys Poisson's equation (eq.\ref{Poisson}), once the boundary conditions are satisfied, the field created by all charges (real and image) is the field obtained in the actual configuration outside the conductors. It is important to state that in the fictitious configuration the charges outside the conductors must be in the same place as in the real distribution, or else one would find a solution to Poisson's equation with a different $\rho({\bf x})$: all image charges must be placed in the space originally occupied by the conductors.

The method can then provide the {\it force} acting on a charge in the presence of conductors. However, if one tries to naively use the image method to compute the electrostatic energy of that configuration, an incorrect result will be found, namely, one arrives at twice the correct energy, as Griffiths \cite{GriffithsBook} and Franklin \cite{JFranklin} have shown for particular cases. In simple situations as the point charge-plane wall setup, it is rather easy to realize that the real configuration has half the energy of the fictitious one, but in more involved and less symmetrical geometries this is no longer an obvious issue. In the case of a charge in front of a sphere one could lucidly expect to find a prefactor depending on the sphere radius which would only tend to $1/2$ if the radius tended to infinity (reobtaining the charge-wall result).

Our goal is to show that, whatever the shape and amount of perfectly conducting bodies near a point charge, the electrostatic energy of the system is $1/2$ times the Coulombic energy of the interaction between the real charge and each image charge of the problem. We shall then generalize that result for more than one source charge. 

Our article is organized as follows: we begin in Section 2 with the familiar problem of a charge and a conducting wall, and also comment on the case of a wedge. We then proceed, in Section 3, to a less symmetrical geometry, that of a conducting sphere. Section 4 is dedicated to the theorem demonstration in the general case, and we leave the last section for conclusion and remarks.

\section{Usual case: point charge and conducting wall}

In order to state the problem and emphasize the important point in
the clearest way we start our discussion by considering in this
section the simplest problem that can be solved by the image method,
namely, to find the electrostatic field of a point charge in the
presence of a perfectly conducting wall. This problem can be found
in many standard textbooks \cite{GriffithsBook,JFranklin,Jackson,Stratton}, so we go directly to the point
here.  For convenience, let us choose our cartesian axes ${\cal
OXYZ}$ such that the region $z<0$ is filled with a perfectly
conducting material and a point charge $q$ is at position $(0,0,z)$,
as shown in Fig. \ref{wall}:
\begin{figure}[ht]
\centering
\includegraphics[width=0.5\columnwidth]{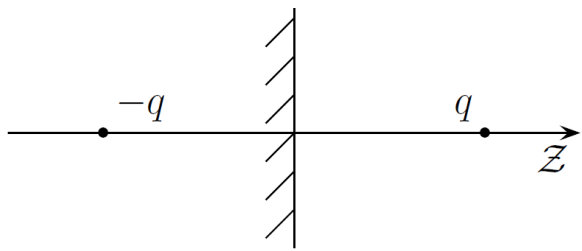}
\caption{Real and image charges for the charge-wall case}
\label{wall}
\end{figure}

\noindent According to the image method, the force exerted on the real charge $q$ by the superficial charge distribution induced on the surface of the conductor is given by
\begin{equation}\label{Forca-q}
{\bf F}_q= -\frac{q^2}{4\pi\epsilon_0}
 \frac{1}{(2z)^2}\, {\hat{\bf z}} \, .
\end{equation}
One could naively think that the electrostatic energy of the
system shown in Fig. \ref{wall} would, too, be given simply by the Coulombic
energy between point charge $q$ and its image, $-q$, namely, $U =
-(q^2/4\pi\epsilon_0)(1/2z)$. However, this is not true, as we can
 easily verify if we take the gradient of $U$,
 \begin{eqnarray}
 -\nabla U = - \hat{\bf z}\frac{\partial}{\partial z}
 \left(-\frac{q^2}{4\pi\epsilon_0}
 \frac{1}{(2z)}\right) =
 -2 \left(\frac{q^2}{4\pi\epsilon_0}
 \frac{1}{(2z)^2}\right) \hat{\bf z}
 \;\ne \;{\bf F}_q\; .
\end{eqnarray}
The correct expression for $U$ has an additional factor of $1/2$ and
can be readily obtained if we start by the very definition of $U$ as
the total external work to bring all the real charges (including the surface distribution) from infinity 
to the final static configuration $C_f$
. The easiest way to compute it is to picture the setup of all charges of the system along their paths from infinity to $C_f$ always consistent with the presence of the conductor. This guarantees that the surface charges move across regions of constant potential, without any work required to bring them. Under
these conditions (in the electrostatic context radiation effects are
negligible), we have
\begin{equation}
U=-W_{ext} = -\int_\infty^{C_f}{\bf F}_q({\bf r}^{\,\prime})\cdot
d{\bf r}^{\,\prime}\, .
\label{work}
\end{equation}
Substituting eq.(\ref{Forca-q}) into the above equation, we
get
\begin{eqnarray}
U=-W_{ext} = \frac{q^2}{4\pi\epsilon_0}\int_\infty^z
\frac{dz^{\,\prime}}{(2z^{\,\prime})^2} =
 -\frac{1}{2}\frac{q^2}{4\pi\epsilon_0}\frac{1}{(2z)}\; .
\end{eqnarray}
In other words, the electrostatic energy of the system formed by point charge $q$ and the conducting region is half the electrostatic energy of a point charge $q$ located at $(0,0,z)$ and a point charge $-q$ located at $(0,0,-z)$. This result is in agreement with (\ref{Forca-q}) as can be readily seen. This kind of discussion can be found in many textbooks, like Griffiths's \cite{GriffithsBook}, among others.

In this simple case, the factor $1/2$ could also be anticipated by
symmetry arguments, as follows. First, recall that $(1/2)\epsilon_0
{\bf E}^2$ is the energy density of the electrostatic field. With
this in mind, we easily see that the energy of a system formed by
the charges $q$ at $(0,0,z)$ and $-q$ at $(0,0,-z)$ (with no
conductor at all) is equally divided between the regions $z>0$ and $z<0$. 
In this calculation, we must, of course, exclude the self-energies of each charge.
%

Symmetry also allows us to deal with the case of a charge near an infinite wedge whose aperture angle equals $\pi/n$ for any positive integer $n$ ($n=1$ corresponding to the plane wall). In these cases, the entire space can be divided into $2n$ sectors with that same angle, one corresponding to the outside of the conductor, the remaining $2n-1$, to the space filled by the conductor. An image charge will be in each sector, except for outside the conductor, where the real charge $q_1$ lies. We shall label the sectors and the pertaining charges with integers, $i=1$ refers to the real charge, $i=(2,...,2n)$, to the image ones.  Symmetry allows us to say that the configuration energy would be $1/2n$ times the Coulombic energy of all $2n$ charges, i.e.,
\begin{equation}
\displaystyle U=\dfrac{1}{2n}\times\sum_{i=1}^{2n}\sum_{j=i+1}^{2n}q_iV_j({\bf r_i}) = \dfrac{1}{2n}\times\sum_{i=1}^{2n}\sum_{j=i+1}^{2n}\dfrac{q_iq_j}{4\pi\epsilon_0r_{ij}} \ ,
\label{1sobre2n}
\end{equation}
where $V_j({\bf r_i})$ is the potential created by charge $q_j$ at the position ${\bf r_i}$ of charge $q_i$ and $r_{ij}$ is the distance between charges $q_i$ and $q_j$. This seems to indicate that the prefactor depends on $n$, i.e., on the wedge angle at hand, but we must pay closer attention to the expression we are comparing the energy with. The interaction energy of the charge-wedge system is $1/2n$ times the energy of the {\it total system composed of $2n$ charges},  while we wish to compare it with the Coulombic interaction between the real charge and each of the images (image-image interactions not being included). This distinction can be very subtle; in the case $n=2$ we wish to compare the actual energy with the energy of the pairs $(q_1,q_2)$, $(q_1,q_3)$, $(q_1,q_4)$, while the double summation on eq.(\ref{1sobre2n}) includes these plus the pairs $(q_2,q_3)$, $(q_2,q_4)$, $(q_3,q_4)$. More generally, the double summation on eq.(\ref{1sobre2n}) comprises \mbox{$n(2n\!\!-\!\!1)$} pairs. If one only counts the interaction between the real charge and each image, one finds $(2n\!-\!1)$ pairs. Moreover, using the fact that odd-numbered charges have the value $q$ and even-numbered ones, the value $-q$, together with the symmetry of the configuration, one can see that
\begin{equation}
\sum_{i=1}^{2n}\sum_{j=i+1}^{2n}\dfrac{q_iq_j}{r_{ij}}=
n \sum_{j=2}^{2n}\dfrac{q_1q_j}{r_{1j}} \ .
\label{fatorn}
\end{equation}
An interested reader may verify eq.(\ref{fatorn}) for any particular value of $n$. We thus conclude that
\begin{equation}
\displaystyle U=\dfrac{1}{2n}\times n\sum_{j=2}^{2n}q_1V_j({\bf r_1}) = \dfrac{1}{2}\sum_{j=2}^{2n}q_1V_j({\bf r_1}) \ ,
\label{fatormeio}
\end{equation}
and once more the energy of the configuration is half of the interaction energy between the real charge and each image.

A natural question then arises: what happens in less symmetric
situations or even in situations where there is no symmetry at all?
 From now on, this answer is our main concern. However, we shall do that in
 two steps. First, we shall consider in the next section another example and work out
 the result explicitly. Then, we shall attack a completely general
 situation of one (or more) charged particle(s) in the vicinity of $N$ grounded or neutral 
 perfect conductors of arbitrary shapes.

 \section{Point charge and a perfectly conducting sphere}

\begin{figure}[ht]
\centering
\includegraphics[width=0.8\columnwidth]{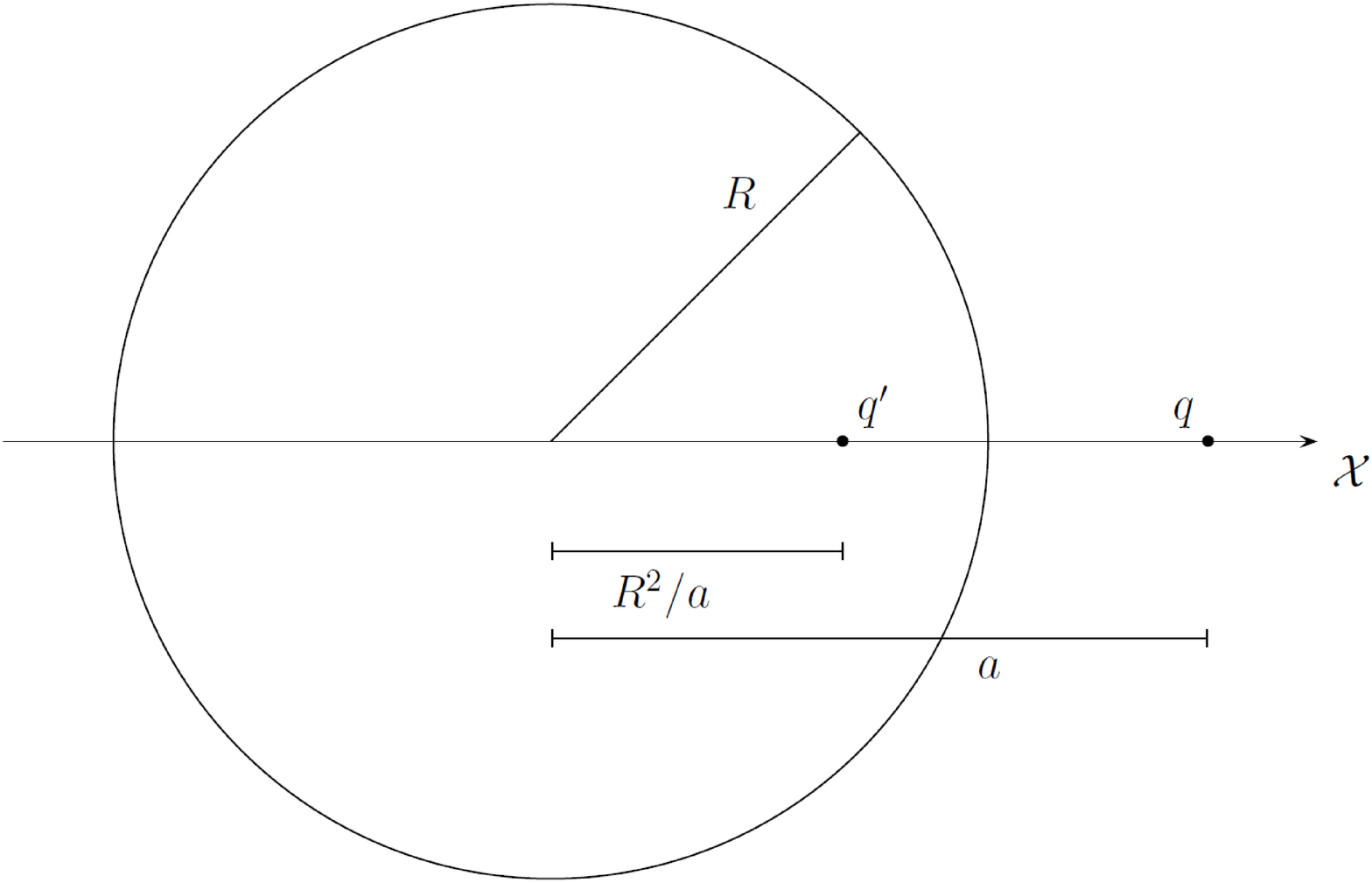}
\caption{Real and image charges for the charge-sphere case. The image charges are $q'=-qR/a$ and $-q'$.}
\label{sphere}
\end{figure}

Let us consider as our next example a point charge $q$ near a perfectly conducting grounded sphere of radius $R$. Suppose the distance from charge $q$ to the center of the sphere is $a$, $a>R$. For simplicity, we choose the axis ${\cal OX}$ with its origin at the center of the sphere so that the position of charge $q$ is given by $(a,0,0)$, as shown in Fig. \ref{sphere}.
It is well known  that the surface charge distribution on the
 sphere is such that the force on $q$ is the same as if there were no sphere at all and a charge $q^{\,\prime} = -(R/a)q$ were located at $(R^2/a,0,0)$ (see Ref. \cite{GriffithsBook}). Therefore, the force exerted by the surface distribution of the sphere on the point charge $q$ is given by
 \begin{equation}\label{Forca-q-Esfera}
 {\bf F}_q = \frac{q(-qR/a)}{4\pi\epsilon_0}\frac{\hat{\bf x}}{(a - R^2/a)^2} 
=  \frac{-q^2R}{4\pi\epsilon_0} \frac{a\,\hat{\bf x}}{(a^2 - R^2)^2}  .
\end{equation}

 According to the previously presented discussion, we do not expect the electrostatic energy of the charge-sphere system shown in Fig. \ref{sphere} to be given by the Coulombic interaction energy between real and image charges. However, we have no reason, a priori, to say that the correct answer is obtained simply by including an additional factor $1/2$ as occurred in the cases discussed in Section 2. It would be natural, though, to expect a factor depending on $R$ and $a$ that, for $R\rightarrow\infty$ with $R-a$ kept constant, reduces to the previous factor $1/2$, since that limit reproduces the charge-wall case.

Let us then perform the explicit calculation using, as before, the very definition of the electrostatic energy of a configuration in light of the comments made before eq.(\ref{work}).  Doing that along the axis
 ${\cal OX}$ and using eq.(\ref{Forca-q-Esfera}), we may write
 \begin{eqnarray}
U = -W_{ext} &=&  \frac{q^2R}{4\pi\epsilon_0} \int_\infty^a \frac{x}{(x^2 - R^2)^2}
dx\cr\cr
 &=& \frac{1}{2} \frac{q(-Rq/a)} {4\pi\epsilon_0(a - R^2/a)} \, ,
 \end{eqnarray}
which is nothing but $1/2$ the Coulombic energy between charge $q$ and its image $q^{\,\prime} = -Rq/a$. At first sight, it seems amazing that the same factor $1/2$ appears. This suggests that this will happen for conductors of general shapes. In fact, this is precisely what happens, as we shall demonstrate in the next section.

\section{General case}
\subsection{One source charge}
\indent We shall now consider one point charge $q$ in the vicinity of a set of $N$ perfect conductors of arbitrary shapes, which can be either neutral or grounded. Let ${\bf x}_0$ be its position in space with respect to some reference frame. Our purpose here is to obtain an expression for the electrostatic energy of this configuration in terms of the Coulombic interaction energy between $q$ and each image charge necessary to solve the problem, which would be the energy necessary to bring in the charge $q$ from infinity with every image held fixed at its final position. For convenience, we shall anticipate the final result in the form of a theorem, namely,
 \begin{quote}
{\it The electrostatic energy of a point charge $q$ near $N$ perfect conductors of arbitrary shapes, each conductor being either neutral or grounded, is half the Coulombic energy between the charge $q$ and each image charge.}
 \end{quote}

We now present a simple demonstration of this theorem. It is convenient to start by the following
expression for the electrostatic energy $U$ of a general charge distribution,
\begin{equation}\label{U-Distribuicao}
U = \frac{1}{2}\int_{\cal R} \rho({\bf x})\, \Phi({\bf x})\, dV\, ,
\end{equation}
where $\rho({\bf x})$ is the charge volumar density at position ${\bf x}$, $\Phi({\bf x})$ is the electrostatic potential at position ${\bf x}$ due to all charge distribution and ${\cal R}$ is a region of space containing all charges. Of course, whenever point charges are present in the distribution, we must subtract from the above expression the corresponding infinite self-energies. It has already been shown in the literature (for instance, in \cite{GriffithsBook}) that this expression is equivalent to computing the external work to bring the charges from infinity.

Since our distribution consists of a point charge $q$ located at
position ${\bf x}_0$ and surface charge distributions on the
conductors, eq.(\ref{U-Distribuicao}) can be written as
\begin{equation}
U = \frac{1}{2}\,q\,\tilde\Phi({\bf x}_0) +
 \frac{1}{2} \sum_{k=1}^N\oint_{S_k}\sigma_k({\bf x})\,\Phi({\bf
 x})\, dA_k\, ,
\label{divided}
\end{equation}
where $\tilde\Phi({\bf x}_0)$ is the potential at position ${\bf
x}_0$ created by all charges of the system except point charge
$q$, $\sigma_k$ describes the  charge distribution on the surface
$S_k$ of the k-th conductor. Writing $\tilde\Phi({\bf x}_0)$
instead of $\Phi({\bf x}_0)$ in the first term of the r.h.s. of the
previous equation is equivalent to subtracting the (infinite)
self-energy of the point charge $q$. Recalling that each surface $S_k$
is an equipotential surface, whose potential we denote by $\Phi_k$, we obtain
\begin{equation}
U =  \frac{1}{2}\,q\,\tilde\Phi({\bf x}_0) +
 \frac{1}{2}\sum_{k=1}^N \Phi_k\oint_{S_k} \sigma_k({\bf x})\,
 dA_k.
\label{prazerar}
\end{equation}
Since each conductor is either grounded, from which $\Phi_k=0$, or neutral, from which the surface integral is zero
, the last term on the r.h.s. of eq.(\ref{prazerar}) always vanishes, and
\begin{equation}
 U =  \frac{1}{2}\,q\,\tilde\Phi ({\bf x}_0) \, 	
	\label{eq:}
\end{equation}

Now, all we need to do is to invoke the image method to finish our
demonstration. Image charges are, by definition, imaginary charges situated in the nonphysical regions (inside the
conductors) that create at any point of the physical region
(outside the conductors) the same field as created by all 
surface distributions of all conductors. Hence, $\tilde\Phi({\bf
x}_0)$ is precisely the electrostatic potential at position ${\bf
x}_0$ due to all image charges, so that we can write symbolically,
\begin{equation}
U = \frac{1}{2}\,q\,\tilde\Phi({\bf x}_0) = \frac{1}{2}\,
U\Bigl(q;\{\mbox{images}\}\Bigr)\, ,
\end{equation}
where $U\Bigl(q;\{\mbox{images}\}\Bigr)$ means all Coulombic interactions
between point charge $q$ and each image. This completes the
demonstration for one point charge $q$.

\subsection{Many source charges}

The result can be further generalized to accommodate the presence of more source charges. We shall now consider a set of $M$ point charges $q_1$, $q_2$, ..., $q_M$ in the vicinity of a set of $N$ perfect conductors of arbitrary shapes, which can be either neutral or grounded. Let ${\bf x}_i$ be the position of charge $q_i$ in space with respect to some reference frame. The expression for the electrostatic energy of this configuration shall also include terms due to interaction of the real charges. Anticipating the final result once more:
 \begin{quote}
{\it The electrostatic energy of a set of $M$ point charges $q_1,q_2,...,q_M$ near $N$ perfect conductors of arbitrary shapes, each conductor being either neutral or grounded, is the Coulombic interaction energy between the real point charges plus half the sum, from $i\!=\!1$ to $i\!=\!M$, of the Coulombic energies between charge $q_i$ and each image charge.}
 \end{quote}

Let us demonstrate this second theorem. We start once again by eq.(\ref{U-Distribuicao}), 
\begin{equation}
U = \frac{1}{2}\int_{\cal R} \rho({\bf x})\, \Phi({\bf x})\, dV\, .
\label{U-Distribuicao2}\end{equation}
We can write the expression separating $\rho({\bf x})$ in the contributions due to each point charge $q_i$ located at position ${\bf x}_i$ and due to the surface charge distributions on the conductors,
\begin{equation}
U = \frac{1}{2}\sum_{i=1}^M\,q_i\,\tilde\Phi_i({\bf x}_i) +
 \sum_{k=1}^N \frac{1}{2}\oint_{S_k}\sigma_k({\bf x})\,\Phi({\bf
 x})\, dA_k\, .
\end{equation}
While $\sigma_k$ describes the charge distribution on the surface
$S_k$ of the k-th conductor, we subtract the self-energy of point charge $q_i$ by substituting the potential $\Phi({\bf x}_i)$ by $\tilde\Phi_i({\bf x}_i)$, the potential at position ${\bf x}_i$ created by all charges of the system {\it except} point charge $q_i$ itself. The second term in the r.h.s. of this equation vanishes analogously to how the second term on the r.h.s. of eq.(\ref{divided}) does. We obtain 
\begin{equation}
 U =   \frac{1}{2}\sum_{i=1}^M\,q_i\,\tilde\Phi_i({\bf x}_i) 	
	\label{somaqi}
\end{equation}
The potential $\tilde\Phi_i$ can be split in the potential due to the other point charges $q_j$  and the potential due to all surface charges together, $\Phi_{surf}$,
\begin{equation}
	 U =   \frac{1}{2}\sum_{i=1}^M\sum_{j=1 \atop j\neq i}^M\dfrac{q_i\,q_j}{4\pi\epsilon_0|{\bf x}_i-{\bf x}_j|} +  \frac{1}{2}\sum_{i=1}^M\,q_i\,\Phi_{surf}({\bf x}_i) 	 \ .
	\label{separ}
\end{equation}
The first term can be readily recognized as the Coulombic energy between the source charges, as can be seen in the literature \cite{GriffithsBook} (we remember that each pair is counted twice in that double summation). Invoking the image method as before, we can state that $\Phi_{surf}({\bf x}_i)$ is  the electrostatic potential at position ${\bf x}_i$ due to all image charges, allowing us to write, symbolically,
\begin{equation}
U = U\Bigl(\{\mbox{all source charges}\} \Bigr) + \frac{1}{2} \sum_{i=1}^M \,
U\Bigl(q_i;\{\mbox{images}\}\Bigr)\,  ,
\label{final}
\end{equation}
thus completing our most general demonstration.

Eq.(\ref{final}) can be interpreted in terms of pairwise Coulombic energies as follows: the energy of each pair composed of two real charges enters the expression integrally; pairs that comprise a real charge and an image one take on a factor one half; pairs of image charges are not present in eq.(\ref{final}).

\section{Conclusions and final remarks}

We have calculated the electrostatic energy of systems composed of a point charge and conductors of various geometries. We started with the simple case of a plane wall and obtained as a result one half of the Coulombic energy between real and image charges, which could be easily understood considering the fields' energy density and the symmetry of the problem. We then indicated, solely on symmetry arguments, that the energy of a system composed of a charge and a wedge of aperture angle $\pi/n$ ($n$ a positive integer) should also be one half of the Coulombic energy between the charge and every image. We then moved on to a less symmetrical geometry, a spherical one, and there, too, we found the same factor $1/2$ when comparing the energy of the actual system to the pairwise Coulombic energy between the real charge and each image. We then proved the theorem that this same factor one half arises in every problem of a point charge in the presence of conductors that is solvable by the image method, whatever its geometry may be. The argument was completed by generalizing the theorem for the case of more than one source charge.

In many situations involving more than one source charge, the interest lies on the interaction energy between the sources and the conductors, but not on the self-energy of the sources. For those cases eq.(\ref{final}) is best written in the form
\begin{equation}
U - U\Bigl(\{\mbox{all source charges}\} \Bigr) = \frac{1}{2} \sum_{i=1}^M \,
U\Bigl(q_i;\{\mbox{images}\}\Bigr)\,  ,
\label{final2}
\end{equation}
which suits multipole sources especially well.

Although our discussion was in the classical context, it may serve as a useful guideline for quantum problems involving dispersive forces between polarizable atoms/molecules and conducting bodies. 

\section*{Acknowledgments}

The authors wish to thank  P.A.Maia Neto, A.Ten\'orio, M.V.Cougo-Pinto, I. Waga and V. Miranda for enlightening discussions, as well as CNPq (Brazil's National Research Council) and Faperj (Research Support Foundation of the State of Rio de Janeiro) for partial financial support.

\end{document}